# High rate, long-distance quantum key distribution over 250km of ultra low loss fibres

D Stucki[1], N Walenta[1], F Vannel[1], R T Thew[1], N Gisin[1], H Zbinden[1,3],
S Gray[2], C R Towery[2] and S Ten[2]
[1] : Group of Applied Physics, University of Geneva, 1211 Geneva 4, Switzerland,
e-mail: hugo.zbinden@unige.ch
[2] : Corning Incorporated, Corning, NY 14831, USA, GrayS@Corning.com

**Abstract**   We present a fully automated quantum key distribution prototype running at 625 MHz clock rate. Taking advantage of ultra low loss (ULL) fibres and low-noise superconducting detectors, we can distribute 6,000 secret bits per second over 100 km and 15 bits per second over 250km.

## Introduction

Quantum Key Distribution (QKD) [1] could well be the second commercial success of quantum physics at the individual quanta level after that of Quantum Random Number Generators [2] (QRNGs). Despite important progress in recent years [3-8], QKD's primary challenge is still to achieve higher bit rates over longer distances. In practice, these should be averaged secret bit rates after distillation and not peak raw rates.

In order to progress towards this challenge one needs:

- to develop new QKD protocols that go beyond the historical BB84 protocol and are especially designed for quantum communication over optical fibre networks,

- to optimize the single-photon detectors, as this is a major limiting factor,

- to use low loss fibres, as the channel loss will ultimately limit the achievable rate of future point-to-point quantum communication.

## 1. QKD protocol and prototype

As an efficient QKD protocol well suited for fibre-based quantum communication we use the Coherent One Way (COW) protocol [3,4]. The system, inspired by classical optical communication, is functional and features low inherent loss: Alice, the transmitter, sends full and empty pulses, Bob, the receiver, temporally distinguishes them with the help of his detector. There are, however, three important differences with classical systems:

- Alice's non-empty pulses are very weak, they contain a mean photon number of 0.5. Consequently, the two kinds of pulses have a common vacuum component and thus do not correspond to orthogonal quantum states.

- As most of Alice's non-empty pulses cannot be detected by Bob, the bits are encoded in pairs of pulses, one empty and one non-empty; the bit value is defined by the position of the non-empty pulse: first=0, second=1.

- For true quantum communication Alice and Bob have to verify the coherence (i.e. perform measurements in a basis conjugate to the data-basis); this is done by Alice sending all pulses with a common phase reference and Bob randomly selecting a small fraction of pulses, not used as data, to send to an interferometer. This coherence is measured between adjacent qubits. This means that an eavesdropper may not individually act on qubits by either removing one photon out of pulses with multiple photons, or by blocking pulses with only 1 photon, without disturbing the system and being detected. Therefore the system is resistant to photon number splitting attacks. To avoid coherent attacks on two pulses across the bit separation, we also send decoy sequences [3,4].

Figure 1 shows a schematic of the COW protocol. The QKD prototype is built around field programmable gate arrays (FPGAs) (Virtex II Pro) and embedded computers for Alice and Bob. The Alice-Bob system controls: an intensity modulator shaping 300 ps pulses out of a cw beam emitted by a DFB laser; all communication between Alice and Bob; Bob's detectors; the secret key distillation; Alice's random number generator (4 Mb/s of QRNG, expanded to >312.5 Mb/s); and the fast electronics. The initialisation and auto alignment procedures involve synchronisation of Alice and Bob's local clocks, optimal timing of the detection window for the detectors and adjustment of the phase between adjacent pulses to the passively stabilized interferometer by tuning the laser wavelength. The final secret bits are then stored, ready to be used, e.g. as in the SECOQC demonstration [9]. More system details can be found in reference [4].

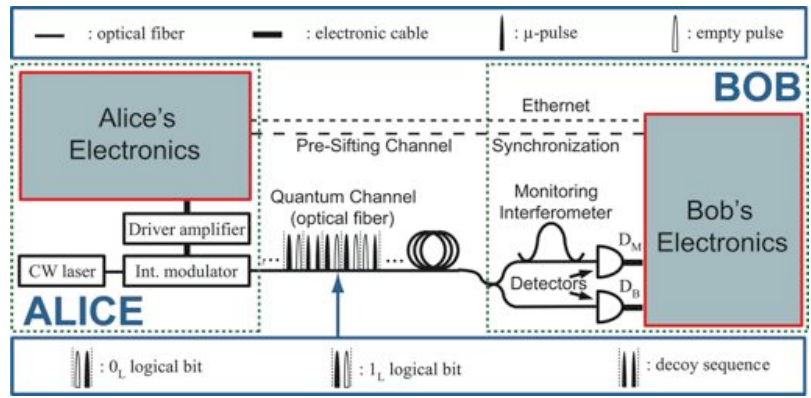

**Figure 1.** Schematic of the Coherent One Way (COW) QKD protocol and its implementation

## 2. SSPD detectors

In order to maximise the potential of the COW prototype we employed two fibre-coupled superconducting single photon detectors (SSPDs) based on meandered NbN nano-wires (figure 2a) [10]. Despite normally being cooled in liquid Helium, SSPDs show great promise for long distance QKD as they have a very low dark noise and the potential for high quantum efficiency. As part of the EU project SINPHONIA [11] we took a step towards their more practical integration and implemented these detectors in a cryogen-free, closed cycle, cryostat system (figure 2b).

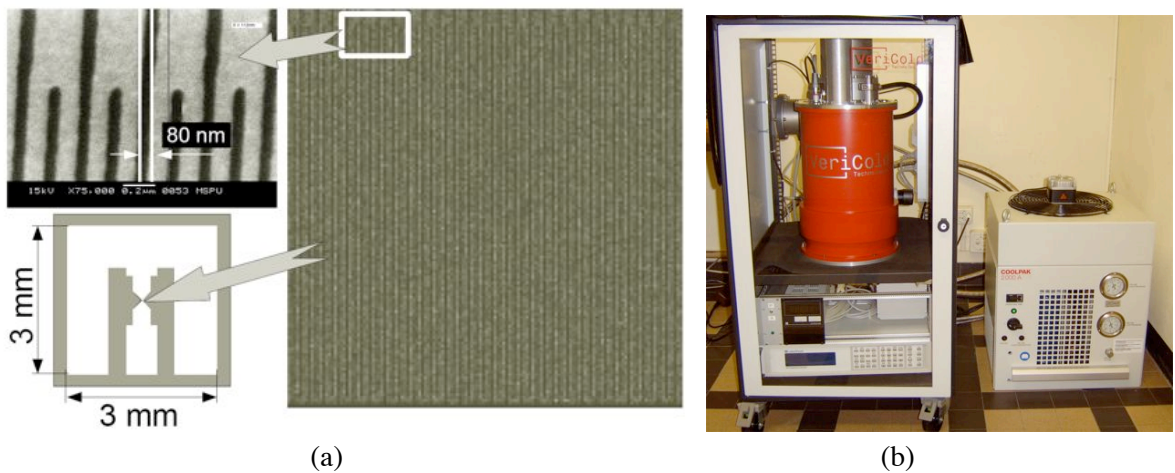

**Figure 2.** (a) Detector contact pad design and SEM image of the superconducting meander covering a 10 $\mu$m x 10 $\mu$m area. The strip width is 100-120 nm with spacings of 80-100 nm respectively. (b) Closed cycle pulse tube cryostat

The quantum efficiency of the SSPDs depends considerably on the cryostat temperature, on the bias current applied and, because of their meander-structure, on the polarisation state of the incident photons [12]. It increases linearly with the applied bias current, while the corresponding dark count noise increases exponentially. Figure 3 shows the dark count rate as a function of the quantum efficiency of the better detector at 2.5 K, when the bias current is scanned. In our experiment, over 250 km, we used this detector for the data line, with a bias current of 26.5 $\mu$A, leading to quantum efficiency of 2.65% at a very low noise rate of 5 Hz. For all measurements we optimized the input polarisation at the beginning of the key exchange. As changes in the environment temperature can affect the input polarization state, we also have to accept variations in the detection efficiency during long-term key exchanges (up to a factor 2). For the monitoring line we used the SSPD, which had a slightly lower detection efficiency at comparable dark count rates.

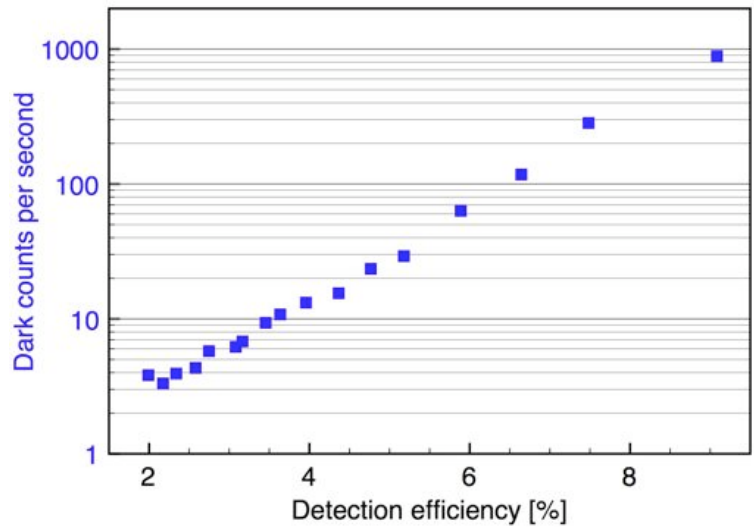

**Figure 3.** Dark count rate vs detection efficiency at 1550 nm and 2.5 K.

### 3. The ultra loss fibres

A major limitation for all quantum communication is fibre loss. To this end, for this QKD experiment we have used a special low-attenuation ITU-T G.652 standard compliant fibre: Corning® SMF-28® ULL fibre [13]. This fibre has a Ge-free silica core (sometimes referred to as pure silica core fibres) with a low Rayleigh scattering coefficient [14]. The effective area and dispersion at 1550 nm are also comparable with standard installed fibre with about 85 $\mu$m$^2$ and 16.1 ps/(nm-km), respectively. Typically these low attenuation pure silica core fibres comply with the ITU-T G.654 standard, and are used in repeater-less submarine transmission systems where the lowest possible attenuation is highly desirable. For terrestrial applications where backwards compatibility with existing fibre plant can be an issue, the SMF-28® ULL fibre offers the advantage of a more straightforward integration into the installed network. However, it is the potential for very low attenuation fibres in terrestrial links that is interesting for QKD. The average fibre attenuation (without splices) is 0.164 dB/km at 1550 nm. Our 250 km link has a total loss 42.6 dB. This is equivalent to 213 km of standard fibre with attenuation of 0.2 dB/km.

### 4. Results

We performed key exchanges in the laboratory over fibre lengths ranging from 100 km to 250 km with secret bit rates from 6 kbits/s to 15 bits/s and Quantum Bit Error Rates (QBER's) from 0.85 to 1.9%, respectively. Figure 4 illustrates the operation of the monitoring line for a 100 km run. In an initial step the laser wavelength is scanned in order to determine the visibility of the interference fringes and the optimal point of operation. After the exchange is started, the counts on the monitor detector are maintained at a minimum value by adjusting the wavelength. The registered counts and the corresponding fringe visibility give us Eve's potential information, which has to be taken into account during privacy amplification. The privacy amplification is implemented using hashing functions based on Toeplitz matrices [15].

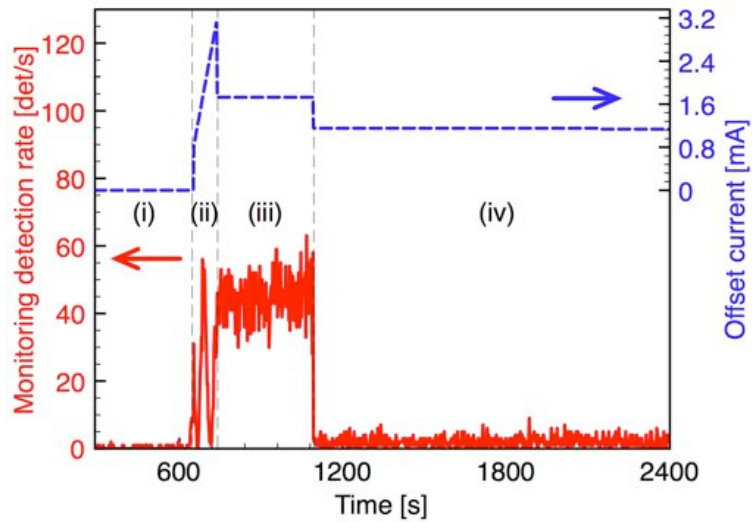

**Figure 4.** Detection rates at the monitor detector (red full line) as a function of time and depending on the wavelength of the laser offset current (blue dashed line). From left to the right, we see the noise measurement (i), scanning of the laser wavelength leading to interference fringes (ii), constant wavelength in order to determine the maximum value (constructive interference) (iii), and wavelength tuned for the minimum (destructive interference) during the key exchange (iv). From (iii) and (iv) we calculate the visibility. In this measurement the visibility was always higher than 92% over two hours.

Figure 5 presents the results obtained over 250 km of SMF-28® ULL fibre with our SSPDs. The horizontal scale denotes the time; at time 0 the prototype starts with an automatic adjustment of the electronic delays and initial alignment of the interferometer. The secret bit distillation then starts only when the raw key buffer is full ($2^{15}$ bits). The error correction is realized with the Cascade algorithm [16]. A Wegman-Carter type scheme, implementing universal hashing functions [17], is used for authentification. Hence, for some initial time no key is generated. Albeit for some minor fluctuations, the system is then stable over hours, producing, on average, more than 15 secret bits per second. Note that at this distance the counts on the monitor detector were too low to correctly determine the visibility and to allow continuous alignment of the laser wavelength. Therefore the security for the key exchange at this extreme distance is questionable. Note that the fibres used for synchronisation/presifting and for classical communications were shorter than the quantum channel (100 km and 25 km, respectively), in order to avoid the need for amplification of the classical signals.

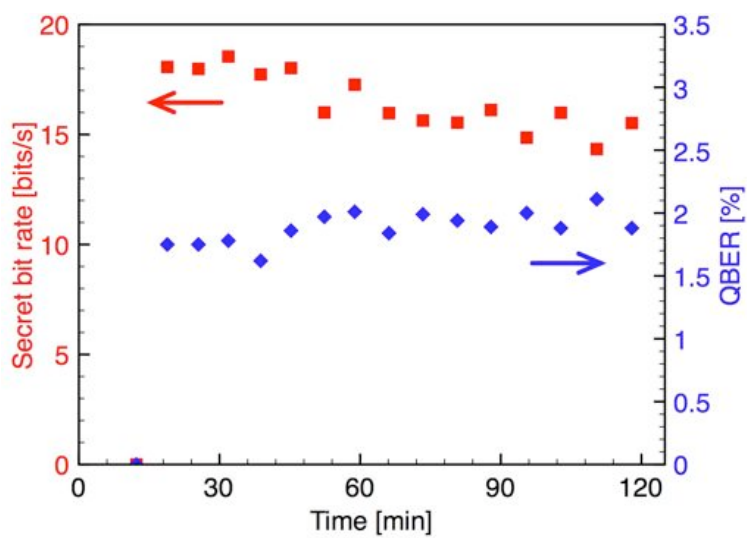

**Figure 5.** Mean secret bit rate per second over 250 km of ULL fibre (red square) and corresponding QBER (blue diamond) as a function of time from the start of the key exchange. Each point is the result of the distillation from $2^{15}$ bits of raw key leading to approx. 7000 secret bits per block.

Figure 6 summarizes several QKDs over distances ranging from 100 km to 250 km. The secret bit rate over 100 km (6 kbit/s) is close to the actual record rate, which was obtained with a laboratory system without synchronisation at a distance [18]. For the last run at 250 km, the temperature of the detector's cryostat was particularly low and as such the data detector was operating with a lower noise and higher quantum efficiency for the same bias current. For this reason, the QBER didn't continue to increase with respect to the other measurements and as a consequence the secret bit rate didn't show the characteristic drop at the end of the range. Note that each point corresponds to the bit rate averaged over 10 minutes or more.

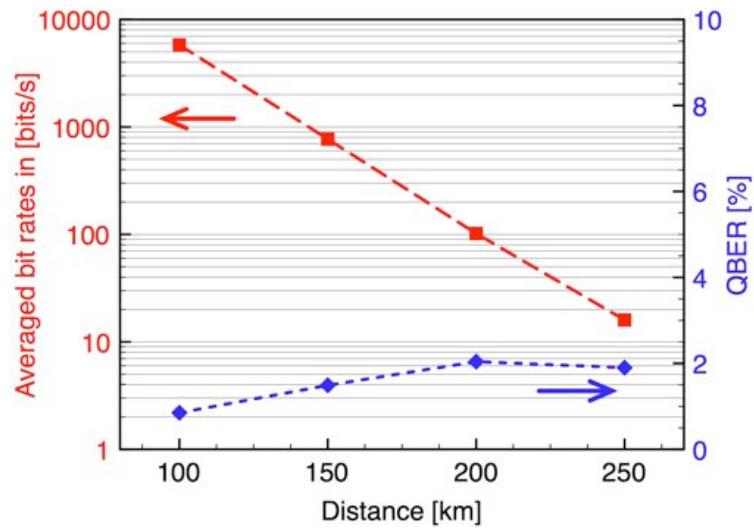

**Figure 6.** Averaged secret bit rates (red squares) and QBER (blue diamonds) for a range of SMF-28® ULL fibre lengths

## 5. Conclusions

The performance of this COW QKD prototype is the product of a large mix of competencies, from theoretical physics (for the security analysis of protocols) to telecom engineers and electronic and software specialists. Moreover, taking advantage of recent progress for SSPDs and optical fibre technology, we have demonstrated a quantum key exchange over a record distance of 250 km of optical fibre with above 15 bit per second. The target of distributing quantum keys over the inter-city distances of up to 300 km with meaningful secret key rates is in sight.

## Acknowledgements

The Geneva group's work was supported, in part, by the EU projects SECOQC and SINPHONIA and by the Swiss NCCR Quantum Photonics. We would like to thank C. Barreiro and P. Eraerds for technical assistance and helpful discussions.